# Physics-informed reinforcement learning for sample-efficient optimization of freeform nanophotonic devices

Chaejin Park[1,2,†], Sanmun Kim[1,†], Anthony W. Jung[2,†], Juho Park[1], Dongjin Seo[1,3], Yongha Kim[2], Chanhyung Park[1], Chan Y. Park[2*], and Min Seok Jang[1*]

[1]School of Electrical Engineering, Korea Advanced Institute of Science and Technology, Daejeon 34141, Republic of Korea

[2]KC Machine Learning Lab, Seoul 06181, Republic of Korea

[3]AI Team, Glorang Inc., Seoul 06140, Republic of Korea

*Corresponding authors. Email: chan.y.park@kc-ml2.com, jang.minseok@kaist.ac.kr

†These authors contributed equally to this work

**Abstract**

In the field of optics, precise control of light with arbitrary spatial resolution has long been a sought-after goal. Freeform nanophotonic devices are critical building blocks for achieving this goal, as they provide access to a design potential that could hardly be achieved by conventional fixed-shape devices. However, finding an optimal device structure in the vast combinatorial design space that scales exponentially with the number of freeform design parameters has been an enormous challenge. In this study, we propose physics-informed reinforcement learning (PIRL) as an optimization method for freeform nanophotonic devices, which combines the adjoint-based method with reinforcement learning to enhance the sample efficiency of the optimization algorithm and overcome the issue of local minima. To illustrate these advantages of PIRL over other conventional optimization algorithms, we design a family of one-dimensional metasurface beam deflectors using PIRL, obtaining more performant devices. We also explore the transfer learning capability of PIRL that further improves sample efficiency and demonstrate how the minimum feature size of the design can be enforced in PIRL through reward engineering. With its high sample efficiency, robustness, and ability to seamlessly incorporate practical device design constraints, our method offers a promising approach to highly combinatorial freeform device optimization in various physical domains.

Nanophotonic devices, having carefully designed arrangements of subwavelength elements that strongly interact with incident light waves, enable precise control of the amplitude, phase, and polarization of light at microscopic scales with unmatched spatial resolution, allowing for highly efficient thin-film solar cells [1], optical information processing and computing [2,3], ultrathin lenses beyond the conventional limits [4-7], and dynamic modulation of complex field amplitude [8-10]. The increasing demand for high-performance, multifunctional nanophotonic devices requires a design method that yields more performant devices than conventional fixed-shape design methods, such as a freeform design approach, which does not impose constraints on the shape or topology of the device to explore potential design candidates that were previously unattainable [11,12]. However, due to the large number of design parameters involved, the computational load of electromagnetic simulation to generate the sample devices for structural optimization are significantly heavier when adopting a freeform approach. The adjoint-based method provides a route to handle such highly combinatorial design problems thanks to its high sample efficiency [13-15], but it is essentially a local optimization algorithm. Conventional population-based heuristics, which have been popularly used for global structural optimization of photonic devices [16-18], become inefficient when dealing with a large number of degrees of freedom (DOF) [11]. This calls for an alternative method for sample-efficient global optimization of nanophotonic devices, and machine learning can be a promising candidate.

Developments in machine learning (ML) techniques have revolutionized the field of photonic device design. Recent studies have verified the capability of neural networks to approximate the relationship between a device's structure and its optical response [19-23]. Additionally, generative models have been proposed to address inverse design problems with high degrees of freedom (DOF) [24-26]. Reinforcement learning (RL) [27], another branch of ML, is recognized to be a competitive approach to solving combinatorial problems [28-30] that have large

DOF. RL has achieved numerous breakthroughs in various problems of combinatorial nature, including the game of Go [31] and the AI accelerator chip design [32], and has also been successfully employed in designing optical metasurfaces [33,34]. However, the requirement for a large number of training samples in ML-based methods raises concerns about the effectiveness of utilizing neural networks in photonic device design, given the substantial computational cost of electromagnetic simulation associated with device sample acquisition. But the very fact that there is an underlying physics can ease the requirement of a large number of training samples by harmoniously combining physics and machine learning.

The practice of incorporating the physics of a system into a neural network to enhance the sample efficiency of machine learning has been investigated in various domains of physical science. For neural networks aimed at predicting physical quantities, such as electromagnetic fields [35], fluid flow [36], and quantum mechanical wavefunctions [37], the governing physical equations of each system can be utilized during the training stage to ensure that the predictions align with the laws of physics. By incorporating this physics-informed approach, neural networks have demonstrated the ability to provide accurate predictions even in the absence of numerically simulated samples [38]. In the field of photonics, physics-informed neural networks have also been employed for device design optimization and inverse design [38,39]. However, there have been few reported progresses in incorporating physical information into RL methods.

In this work, we introduce physics-informed reinforcement learning (PIRL), which combines the physical information from the adjoint-based method with deep RL. With PIRL, we address the optimization problem of a one-dimensional freeform metasurface beam deflector with a combinatorial design space as large as $\sim 10^{74}$. By pre-training an RL agent using the physical information, PIRL demonstrates significantly higher sample efficiency compared to the previously developed RL approach [33]. Moreover, when compared to previous studies on the same design problem, the optimal devices discovered through PIRL generally exhibit superior performance with reduced variance in terms of the same figure of merit. We also demonstrate that the sample efficiency of PIRL can be further enhanced by employing the transfer learning method [40] to the RL agent network from one design problem to similar problems. Finally, we show that practical device design constraints, such as enforcing a minimum feature size for fabrication compatibility, can be seamlessly incorporated into our PIRL framework through simple reward engineering of RL [27].

**METHODS**

Our design objective is to create a one-dimensional silicon metagrating placed on a silica substrate. This metagrating functions as a beam deflector for a normally incident transverse magnetic (TM) polarized plane wave with a wavelength $\lambda$, redirecting the beam to a first-order diffraction angle $\theta$, as illustrated in Figure 1a. The refractive index of silica is set to 1.45, and we use the same index for silicon as in previous publications on the same system [33,39]. The height of the silicon pillar is $h = 325$ nm, and the grating period is $P = \lambda / \sin\theta$, determined by the condition for first-order diffraction. The period is divided into $N = 256$ uniform cells. Each cell of the metagrating can be filled with either air or silicon, and the metagrating structure is represented as a $1 \times 256$ array, $s_t$, where the $i$-th element specifies the material of the $i$-th cell (+1 for Si and -1 for air), as shown in the right panel of Figure 1a. Our goal is to find an optimal structure that achieves the highest possible deflection efficiency, $\eta$, at a given wavelength and angle. We consider a target range of deflection angles and wavelengths, $\lambda = \{900, 1000, 1100\}$ nm and $\theta = \{50, 60, 70\}$ degrees, as adopted from previous studies, for direct comparison [33,39]. We focus on the $N = 256$ case, which entails identifying an optimal structure from approximately $(2^{256}/256) \approx 10^{74}$ possible configurations, excluding degeneracy resulting from cyclic permutation. It is worth noting that the size of the design space in this problem is comparable to the number of atoms in the universe (~$10^{80}$) [41].

The overall procedure of PIRL comprises two stages: a pre-training stage using supervised training and a fine-tuning stage using RL, as depicted in Figure 1b and Figure 1d, respectively. During the pre-training stage, a neural network is trained with the current state of the one-dimensional metagrating as an input, and a proxy of the efficiency gain (*Δη*) associated with flipping each cell from Si to Air or vice versa in the metagrating as a prediction. By leveraging the Lorentz reciprocity [42], the gradient of *η* with respect to the refractive indices of the cells, $n_i$, can be estimated with only two electromagnetic simulations regardless of the number of cells involved, as shown in the top panel of Figure 1b [43]. Using the efficiency gradient, $\partial\eta/\partial n_i$, the efficiency change resulting from flipping the *i*-th cell can be approximated as $\Delta\eta_{\text{approx}} = (\partial\eta/\partial n_i)\,\Delta n$. $\Delta\eta_{\text{approx}}$ is then normalized by its L2 norm for the stability of the training of the neural network and used as an output of the supervised training. The input and output vectors of the adjoint gradient prediction network have the same size *N*, and the *i*-th entry of the output vector is correlated to the *Δη* of the device for flipping the *i*-th cell in the input structure. For the architecture of the neural network, we employ a U-Net [44], which is commonly used as a function approximator in the photonics domain [21,35]. In the U-Net, features are extracted from the input through the encoding network and mapped to the output through the decoding network. Skip connections are utilized between the encoding and decoding layers to preserve spatial information. Additionally, to account for the periodic nature of the deflector, our neural network employs cyclic padding for the convolutional layers. Further details of the network architecture are provided in Figure S1. The network is trained to minimize the mean squared error loss between the predictions and the normalized $\Delta\eta_{\text{approx}}$ calculated from the adjoint-based method. The training dataset consists of 20,000 pairs of structures and adjoint gradients, with the number of training samples chosen to strike a balance between maximizing sample efficiency and achieving higher training accuracy. The prediction error as a function of the training sample size is plotted in Figure S2a, and additional information regarding the configuration of the training dataset is presented in Figure S2b.

The predictions of the pre-trained neural network correlate well with $\Delta\eta_{approx}$ obtained from adjoint gradients, as demonstrated in Figure 1c. It is worth noting that the actual efficiency difference resulting from a flip action, $\Delta\eta_{exact} = \eta_{after\ flip} - \eta_{before\ flip}$, may slightly differ from the gradient-based $\Delta\eta_{approx}$. This discrepancy arises because the refractive index change associated with flipping a cell, $|\Delta n| = |n_{Si} - n_{air}| \approx 2.5$, is substantial (Figure S3). However, because calculating $\Delta\eta_{exact}$ for a device using a finite difference approach would require $N + 1 = 257$ simulations, instead, we have opted to utilize $\Delta\eta_{approx}$, which involves only two simulations and is thus more than two orders of magnitude computationally efficient. Despite the significantly reduced computational cost, $\Delta\eta_{approx}$ and the neural network predictions exhibit a similar trend to $\Delta\eta_{exact}$ as illustrated in Figure 1c.

The pre-trained network is then utilized as the initial weights of the agent's network in the RL stage, as illustrated in Figure 1d. Unlike the pre-trained network with a myopic perspective, the RL agent learns to pursue long-term returns, even if it entails short-term losses, through deep Q-learning [27]. This aspect is crucial for a global optimization method, since relying solely on immediate rewards may lead to convergence to local optima. During the RL stage, the agent explores the design space by iteratively interacting with the environment, which is represented by a rigorous coupled-wave analysis (RCWA) solver [45]. The interaction involves exchanging information such as state, action, and reward. The agent selects an action based on the current state, and the environment provides a reward as a consequence of the action.

In our approach, the state $s_t$ is represented by a vector of length $N$, which corresponds to the metagrating structure as defined in the right panel of Figure 1a. At each step $t$, the transition from state $s_t$ to $s_{t+1}$ occurs through the action $a_t$. The action $a_t$ is defined as flipping the material (silicon and air) in one of the cells. Therefore, the action space is the cell number (1, 2, ..., $N$) that will be flipped. The reward $r_t$ is defined as the change in optical efficiency $\Delta\eta = \eta_{t+1} - \eta_t$ resulting from the action $a_t$. This reward setting allows the RL objective function, which is the sum of sequential rewards, to be equivalent to the final change in optical efficiency after a series of consecutive actions along the trajectory. Introducing the discount factor $\gamma$, the discounted return $G_t$ is defined as Eqn. 1.

$$G_t = \sum_{t=0}^{T} \gamma^i R(s_{t+i}, a_{t+i}) \tag{1}$$

We set $1 > \gamma \geq 0.99$ to ensure that the discounted return provides a sufficiently accurate approximation of the net change in deflection efficiency over the trajectory. By choosing a value of $\gamma$ close to 1, we emphasize the long-term impact of actions on the overall optimization process. This allows the RL agent to prioritize actions that lead to substantial improvements in deflection efficiency, even if they result in temporary reductions along the trajectory.

The trial-and-error process in RL is formally represented as a Markov Decision Process (MDP), described as a time series tuple $(s_t, a_t, r_t, s_{t+1})$. The policy $\pi$, also known as the decision-making function, determines how the agent selects actions $a_t$ given a state $s_t$. The Q-function, denoted as $Q_\pi$, estimates the expected return of taking action $a_t$ at state $s_t$, under the policy $\pi$. The Q-function of the optimal policy is denoted as $Q^*$, which is updated according to the Bellman equation (Eqn. 2) [46]. However, since the exact $Q^*$ cannot be explicitly evaluated due to the huge state space, a neural network is used as a function approximator.

$$Q^* = \mathbb{E}\left[r_{t+1} + \gamma \max_{a'} Q^*(s_{t+1}, a') | s_t = s, a_t = a\right] \quad (2)$$

In this study, we utilize a physics-informed neural network denoted as $Q^\omega(s, a)$ to model the Q-function. This neural network takes a state vector as an input and predicts the Q-value for each action as the output. The $Q^\omega(s, a)$ is physics-informed as it is initialized with the adjoint gradient predicting network at the beginning of the RL process.

During RL, the agent follows the Epsilon-Greedy algorithm [27]. In this algorithm, the agent chooses either a random action with a probability $\epsilon$ (exploration) or the action with the highest Q-value according to $Q^\omega(s, a)$ (exploitation). The exploration probability, $\epsilon$, linearly decreases from 0.99 (exploration-dominant) to 0.01 (exploitation-dominant) during the first half of the RL stage and remains constant at 0.01 during the second half.

Throughout the RL stage, the agent accumulates a trajectory consisting of states, actions, and rewards in the experience replay buffer, which serves as the agent's memory. The agent is trained using randomly selected data from the replay buffer. The weights of the agent's network are updated using the Huber loss [47] and Adam optimizer [48]. The summary of the PIRL algorithm is provided in Table S1,2.

The computation involved in the RL process is parallelized using Ray [49] as depicted in Figure 1e. This parallelization allows for data collection by multiple workers and asynchronous network updates [50]. In this setup, sixteen workers each have their own copy of the central Q-network and interact with their own copy of the environment in parallel, collecting trajectories and storing them in the central experience replay buffer. All the hyperparameters used in the RL stage are provided in Table S6. The total number of electromagnetic simulations performed during the RL stage is set to 200,000, which is comparable to previous work on the same design problem [39].

Each RL stage takes approximately 1.3 hours to complete on a server computer equipped with four Nvidia RTX 3080 GPUs and two Intel Xeon Gold 5220 processors.

**RESULTS/DISCUSSIONS**

The PIRL algorithm generally outperforms other device optimization methods. Figure 2a illustrates the optimization curves of various methods for the representative case of $\lambda = 1100$ nm and $\theta = 60°$. The optimization statistics were collected from ten different executions. The RL-based approaches exhibit a gradual increase in efficiency as the number of simulations increases. In contrast, the greedy algorithm, which selects the cell with the highest efficiency gain ($\Delta\eta$) at each step, quickly converges to a local optimum with high dependence on initial conditions. If the PIRL agent is not trained during the RL stage and instead follows the adjoint gradient learned during the physics-informed pre-training stage, the optimization curve would resemble that of the greedy algorithm since the adjoint gradient predicts immediate rewards. However, by training the agent to approximate the discounted return in Eqn. 1, the agent effectively mimics an infinite-depth greedy algorithm. This leads to slower convergence but with higher terminal efficiency. On the other hand, the optimization curve of the genetic algorithm (GA) shows slower convergence compared to the RL-based methods and does not reach an optimum value within 200,000 simulations. While there is a possibility for the GA to eventually find a better device, its low sample efficiency limits its effectiveness in optimizing devices with high degrees of freedom (DOF).

Among the RL variants, PIRL achieves the highest deflection efficiency with the fastest rate of improvement. Uninformed RL, where the Q-networks are randomly initialized without pre-training, is tested with two different network architectures: U-Net and a fully connected network (FCN). Between the two versions, U-Net outperforms FCN in terms of convergence speed and final $\eta$ value. Despite having a similar number of trainable weights as U-Net, the inherent network architecture of U-Net, which specializes in mapping geometric features from inputs to outputs, likely contributes to its superior performance.

Figure 2b summarizes the performance of various optimization methods, demonstrating that PIRL also outperforms the adjoint-based method and a physics-assisted generative model GLOnet [39] in terms of average and maximum optimized $\eta$ for this specific problem. This trend is consistent across problems with different target conditions, as summarized in Table 1. It is important to note, however, that the results presented in Figure 2b and Table 1 should be taken with a grain of salt, as fine-tuning each algorithm could lead to improved results. A summary of each algorithm can be found in Table S1-5.

**Transfer learning with different target deflection angle conditions**

Transfer learning [40] can enhance the sample efficiency of PIRL. In transfer learning, a neural network trained for a specific wavelength and deflection angle can be utilized for optimizing devices with different wavelength or angle conditions. There are two types of transfer learning applicable to PIRL, which are color-coded in blue and green in Figure 3a. The first type involves transferring a pre-trained network to a different condition, while the second type transfers the fully optimized agent network from one condition to another. Figure 3a also depicts the regular PIRL and uninformed RL without any transfer learning, shown in gray and orange, respectively.

To assess the effectiveness of transfer learning, we compare the deflection efficiencies of the final devices obtained from both transfer learning cases with the outcomes of PIRL and uninformed RL, as presented in Figure 3b. Surprisingly, even when using a pre-trained model with a mismatched pre-training dataset for $\lambda = $ 1100 nm and $\theta = 60°$, transferring it to other angle conditions yields optimization performance similar to that of proper PIRL, which significantly outperforms uninformed RL. These results are remarkable because the state vector, i.e. the configuration of the deflection grating, optimized for one condition often leads to much lower deflection efficiency for a different target condition, as demonstrated in Figure S6.

In contrast, transferring a fully trained RL agent from one condition to another proves to be ineffective and yields results comparable to those of uninformed RL with a randomly initialized Q network. Although the fully trained agent typically starts with better optimization performance, it eventually converges into a low-performance device. In other words, it appears that a pre-trained network exhibits enough flexibility to adapt to a new target condition, while a fully trained network is too rigid to effectively learn new strategies to escape the local optimum from which it starts. Similar behaviors of pre-trained deep neural networks for fine-tuning have been observed in previous studies of meta-learning [51].

**Enforcing the minimum feature size**

In general, optimal devices found using the vanilla PIRL lack fabricability. For example, the device optimized for $\lambda$ =1100 nm and $\theta$ = 60° in Figure 4a can hardly be fabricated even with cutting-edge facilities. This is because the minimum feature size (MFS) of the device is approximately 5 nm, which corresponds to the width of a single cell in the design grid. Simply removing these small features through a Gaussian filter with a half-MFS standard deviation is not a viable solution, as it leads to a catastrophic failure with a drastic drop of 90%p in deflection efficiency, as shown in Figure 4b.

To address these fabricability concerns, we propose a method for enforcing the MFS constraint within the PIRL framework by modifying the reward function. In the MFS-enforced

PIRL, fabricability is incorporated into the reward by subtracting a penalty function $\alpha \Delta B$ from the original reward $\Delta \eta$. Here, $B$ represents the number of pillars or gaps that fall below the MFS limit, and $\alpha$ is a penalty constant that determines the level of enforcement. The value of $\alpha$ is empirically determined by selecting the minimum value that ensures the device satisfies the MFS condition. It's important to note that with this reward setting, the undiscounted return corresponds to the net change in efficiency over the trajectory, assuming the final structure doesn't violate the MFS constraint.

Figure 4c showcases the optimized structure obtained using the MFS-enforced PIRL for $N = 256$ and MFS = 16 cells, with $\alpha$ set to 0.1. It achieves a deflection efficiency of 74.0%, significantly surpassing the optimal structure among the subset of $N = 16$ (Figure 4d). Our method also discovered a device with an efficiency of 85.1% for the problem of $N = 256$ and MFS = 8 cells, outperforming the optimum found for $N = 32$ by 1.3%.

Another approach to enforcing the MFS constraint is by reducing the number of grid cells, $N$, to match the required MFS. However, this coarse grid approach is also undesirable because it confines the design space to a tiny subset of possible outcomes as gaps between features smaller than the minimum feature size can be fabricated in reality. Even the global optimum structure obtained from an exhaustive search within this limited design space has a significantly lower deflection efficiency. For example, the global optimum structure obtained for $N = 16$ (MFS ~ 80 nm), as depicted in Figure 4d, achieves a deflection efficiency of 68.1%. This is more than 5%p lower than the optimal structure found for the same MFS with $N = 256$ in Figure 4c, which is found using MFS-enforced PIRL.

**CONCLUSION**

In this work, we introduce PIRL, a method that integrates physical information from the adjoint-based method into RL for designing highly complex optical devices. By initializing the RL agent's network with the figure-of-merit gradient, we significantly enhance the sample efficiency for optimizing the structure, surpassing the previous work by more than an order of magnitude [33]. To demonstrate the effectiveness of PIRL, we directly compare it with other existing methods, ranging from conventional genetic algorithms to deep generative models [39]. Furthermore, we show that transfer learning can further improve the sample efficiency of PIRL by successfully transferring networks between design problems with different target conditions. Additionally, we address the need for fabrication feasibility by modifying PIRL to enforce a minimum feature size in devices through reward engineering.

This work represents the initial endeavor to incorporate physical information (adjoint gradient) into RL for the design of highly complex optical devices. The developed method can be extended to other techniques as long as the simulation tool allows for the calculation of local gradients of each design element within a limited number of simulations. For instance, automatic differentiation-enabled RCWA tools have the capability to compute local gradients of design elements in a comparable timescale to the adjoint method [45,52-54]. By combining such tools with PIRL, the optimization of devices with intricate figure-of-merit functions becomes feasible. We anticipate that this optimization method will empower RL to address seemingly intractable problems in photonics device design.

**Acknowledgments**

**Funding:** This work was supported by the LX Semicon - KAIST Future Research Center.

**Author contributions:** M.S.J. conceived the presented idea. C.J.P., S.K., and C.Y.P. contributed to the further development of the idea. C.J.P., J.P., and D.S. designed the model and the computational framework. A.W.J. did code parallelization and resolved the technical problems in RL. Y.K. developed the electromagnetic simulation Python package. C.H.P. performed numerical simulations on COMSOL multiphysics. C.J.P. and S.K. carried out simulations and analyzed the data. C.J.P., S.K., and A.W.J. wrote the manuscript with comments and revisions from C.Y.P. and M.S.J.. C.Y.P. and M.S.J. supervised the project.

**Conflict of interests:** The authors declare no competing interests.

**Data and code availability:** The source code is available from the following GitHub repository.

[https://github.com/jLabKAIST/physics-informed-metasurface]

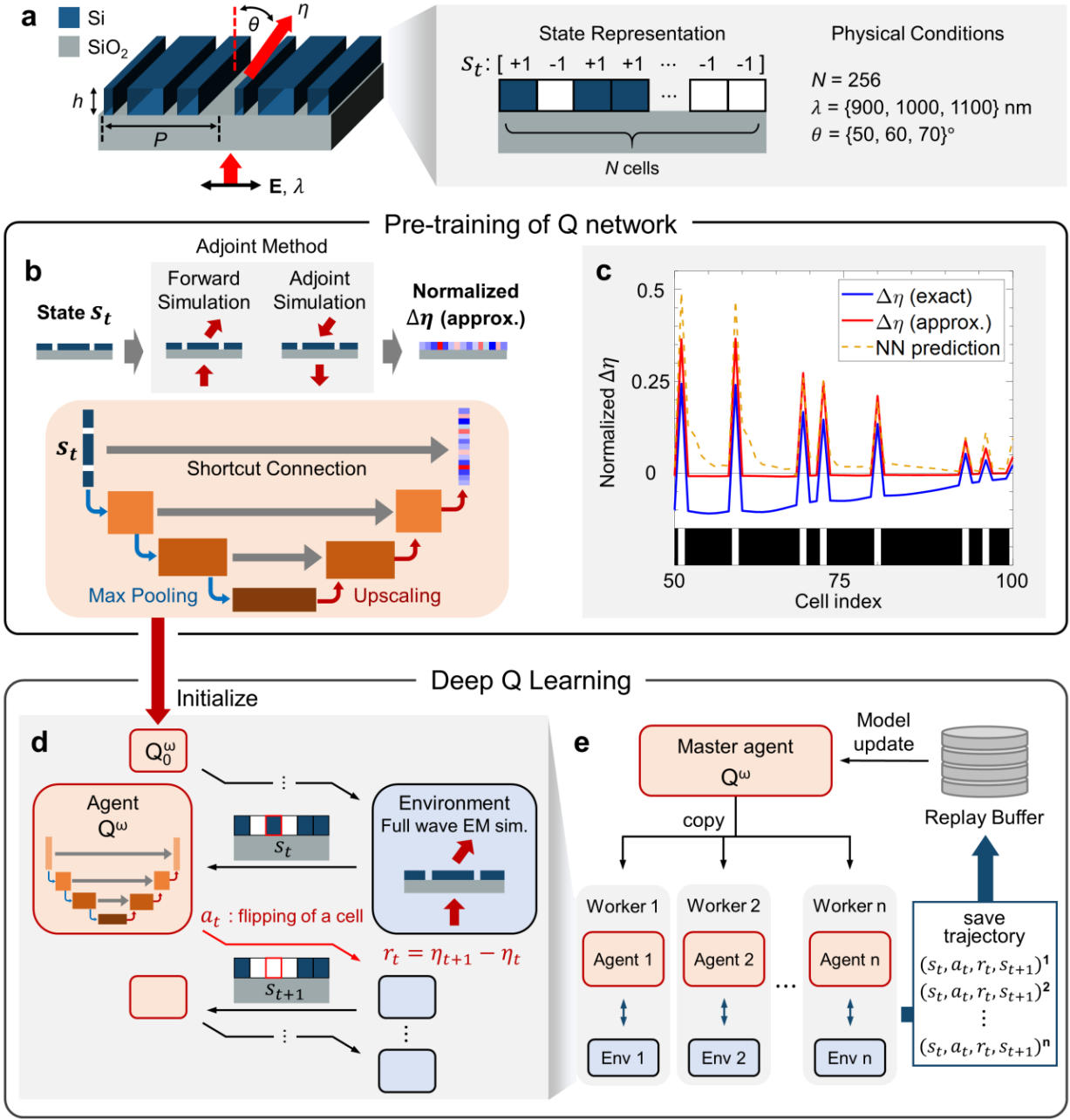

**Fig. 1. Summary of PIRL consisting of a pre-training stage and an RL optimization stage. a,** Schematic diagram illustrating the one-dimensional metagrating and its state representation. The metagrating is composed of silicon pillars on a silicon dioxide substrate. The goal is to maximize the first-order deflection efficiency, $\eta$, for normally incident light with transverse magnetic (TM) polarization. **b,** The physics-informing pre-training stage of PIRL. The U-net shaped agent network, is pre-trained to predict the normalized $\Delta\eta_{approx}$ of a given structure. The samples for network training are generated by the adjoint-based method illustrated in the top panel. **c,** Comparison of $\Delta\eta_{exact}$ (blue), $\Delta\eta_{approx}$ (red), and the prediction result from the pre-trained neural network (yellow dashed line). **d,** Illustration depicting how the agent interacts with the environment in RL. Definitions of state, action, and reward are provided. The pre-trained agent network serves as the initial state of the agent's network. **e,** The parallelized RL stage comprises a master agent $Q^\omega$ and sixteen workers. Each worker has a copy of the agent network obtained from the master agent and independently generates trajectories by interacting with the environment.

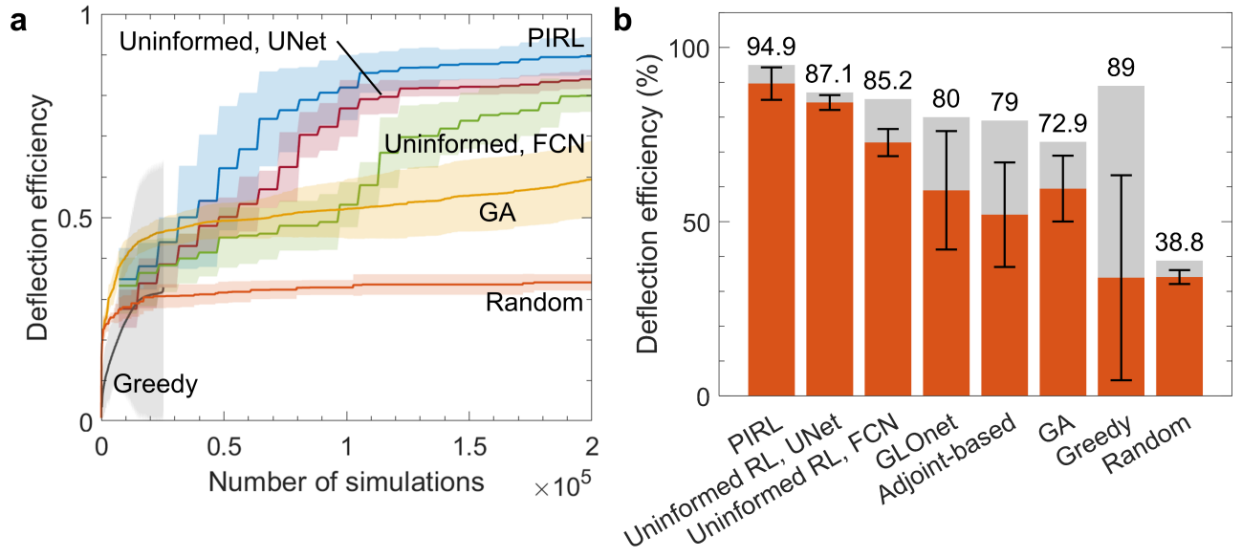

**Fig. 2. a,** Optimization curve showing the performance of PIRL (blue), U-Net based uninformed RL (red), fully connected network (FCN) based uninformed RL (green), genetic algorithm (GA) (yellow), and brute force algorithm (orange) under the target condition $\lambda = 1100$ nm, $\theta = 60°$. The solid line represents the average maximum efficiency, and the shaded area represents the standard deviation over ten executions. **b,** Performance comparison among the optimization algorithms. The orange bar represents the average value of the maximum device efficiency, and the gray bar represents the overall maximum over ten runs. The data for GLOnet was obtained from Jiaqi Jiang et al. [55]. The error bars centered at the average value indicate the standard deviation over ten optimization runs. Additional details, including the average, standard deviation, and the deflection efficiency of the best performing devices, can be found in Table 1. A summarized algorithm table comparing the algorithms can be found in Table S1-5.

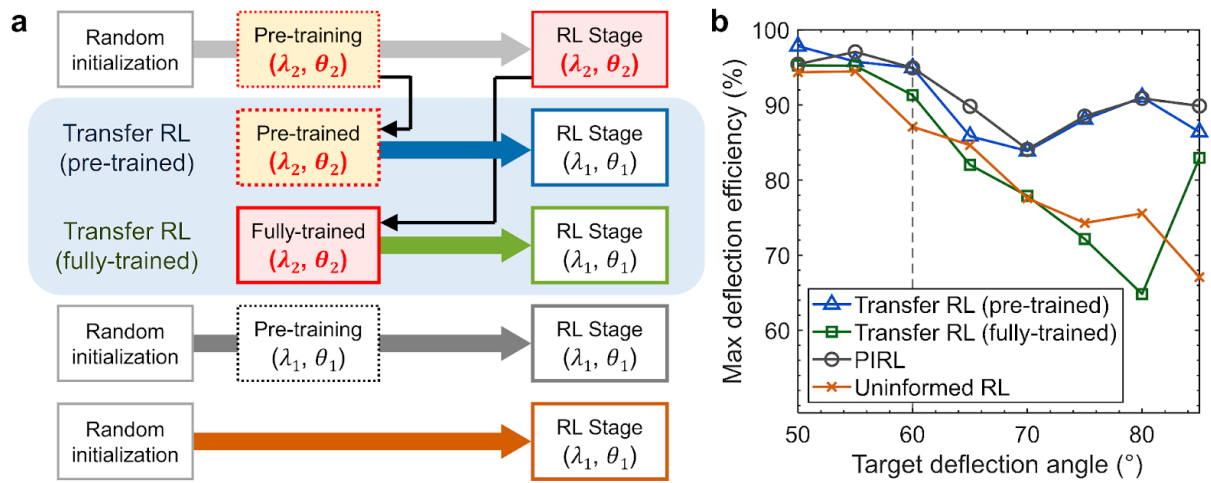

**Fig. 3. a,** Schematic diagram illustrating two different transfer learning processes. In the first case (blue arrow), a neural network trained with the adjoint gradients of condition ($\theta_1$, $\lambda_1$) is used as the initial network for the RL agent optimizing the problem ($\theta_2$, $\lambda_2$). In the second case (green arrow), the neural network that underwent the full PIRL process for condition ($\theta_1$, $\lambda_1$) is used as the initial network for the RL agent optimizing the problem ($\theta_2$, $\lambda_2$). **b,** Maximum deflection efficiency of the device obtained using each transfer learning method. Both PIRL and transfer RL with a pre-trained network outperform transfer RL with a fully-trained network and uninformed RL. The averages and standard deviations can be found in Figure S7.

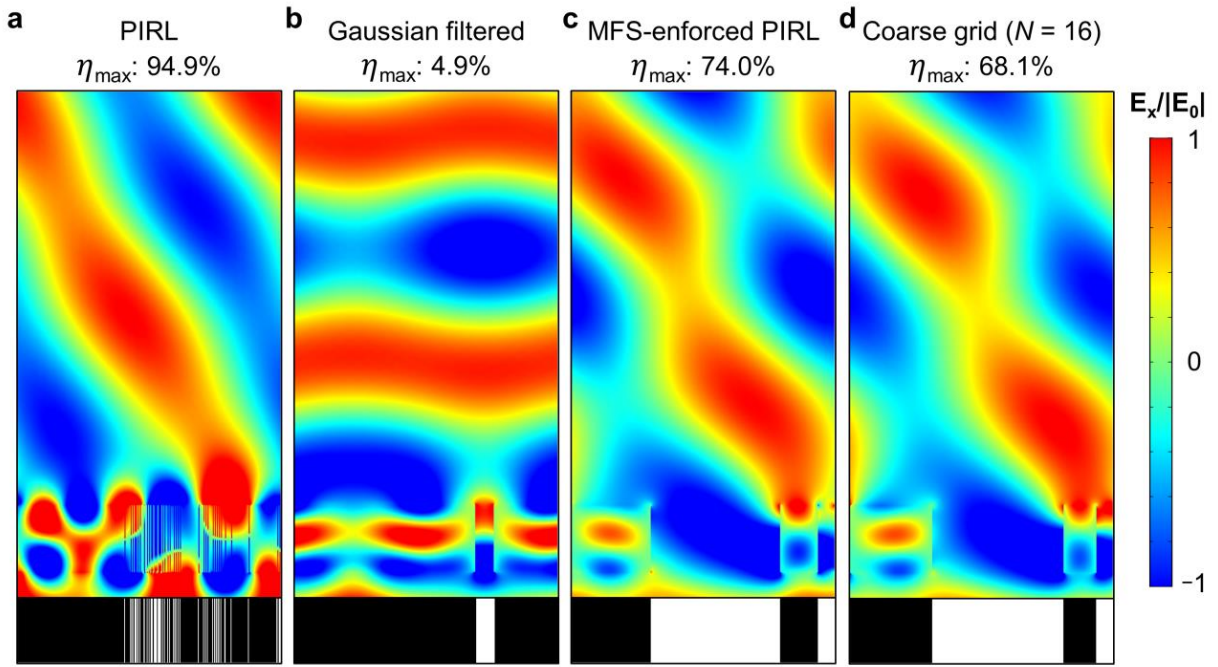

**Fig. 4. Electric field profiles from beam deflectors designed for a target condition of $\lambda$=1100 nm and $\theta$=60°.** All electric field profiles are normalized to the electric field intensity of the incident wave. **a,** Structure of the highest deflection efficiency device found with PIRL along with the corresponding electric field distribution. However, this device cannot be fabricated even with cutting-edge fabrication techniques. **b,** Structure and electric field resulting from the Gaussian-filtered device. The device found with PIRL is filtered using a Gaussian filter with a standard deviation $\sigma$=4 and then binarized. Although the minimum feature size is increased to eighteen, the deflection efficiency drops by 86.45%p. **c,** Electric field profile and device structure of the beam deflector obtained from MFS-enforced PIRL. The smallest feature in the device is a gap of 16 cells. **d,** Electric field profile and device structure of the beam deflector obtained from the resolution-limited device (DOF=16). An exhaustive search was conducted to find the true global optimum.

**Table 1. Maximum, average, and standard deviation of final devices from PIRL, GLOnet, and adjoint-based method for target conditions of $\lambda$ = {900 nm, 1000 nm, 1100 nm} and $\theta$ = {50°, 60°, 70°}.**

| Target condition | | PIRL | | GLOnet [55] | | Adjoint-based method [55] | |
|---|---|---|---|---|---|---|---|
| $\lambda$ | $\theta$ | Max | Mean±Stdev | Max | Mean±Stdev | Max | Mean±Stdev |
| 900 nm | 50° | 97.6 | **96.0±1.3** | **98** | 90±10 | 93 | 64±16 |
| | 60° | **99.7** | **99.5±0.2** | 97 | 73±18 | 93 | 59±18 |
| | 70° | **99.1** | **98.6±0.4** | 98 | 83±14 | 92 | 59±13 |
| 1000 nm | 50° | **97.9** | **91.1±6.2** | 96 | 85±12 | 95 | 55±16 |
| | 60° | **98.7** | **97.4±1.3** | 98 | 85±17 | 92 | 56±14 |
| | 70° | **93.9** | **87.8±4.3** | 93 | 76±18 | 84 | 62±12 |
| 1100 nm | 50° | **95.4** | **93.3±3.5** | 91 | 77±11 | 91 | 49±10 |
| | 60° | **94.9** | **89.6±4.7** | 80 | 59±17 | 79 | 52±15 |
| | 70° | **84.1** | **77.8±3.6** | **84** | 65±14 | **84** | 59±14 |

# Supplementary Materials for

# Physics-informed reinforcement learning for sample-efficient optimization of freeform nanophotonic devices

Chaejin Park[1,2†], Sanmun Kim[1†], Anthony W. Jung[2,†], Juho Park[1], Dongjin Seo[1,3], Yongha Kim[2], Chanhyung Park[1], Chan Y. Park[2*], Min Seok Jang[1*]

[1]School of Electrical Engineering, Korea Advanced Institute of Science and Technology, Daejeon 34141, Republic of Korea

[2]KC Machine Learning Lab, Seoul 06181, Republic of Korea

[3]AI Team, Glorang Inc., Seoul 06140, Republic of Korea

Correspondence to:
* chan.y.park@kc-ml2.com
* jang.minseok@kaist.ac.kr

†These authors contributed equally to this work.

**This PDF file includes:**

    Supplementary Text
    Figs. S1 to S5
    Tables. S1 to S6

## S1. Network architecture

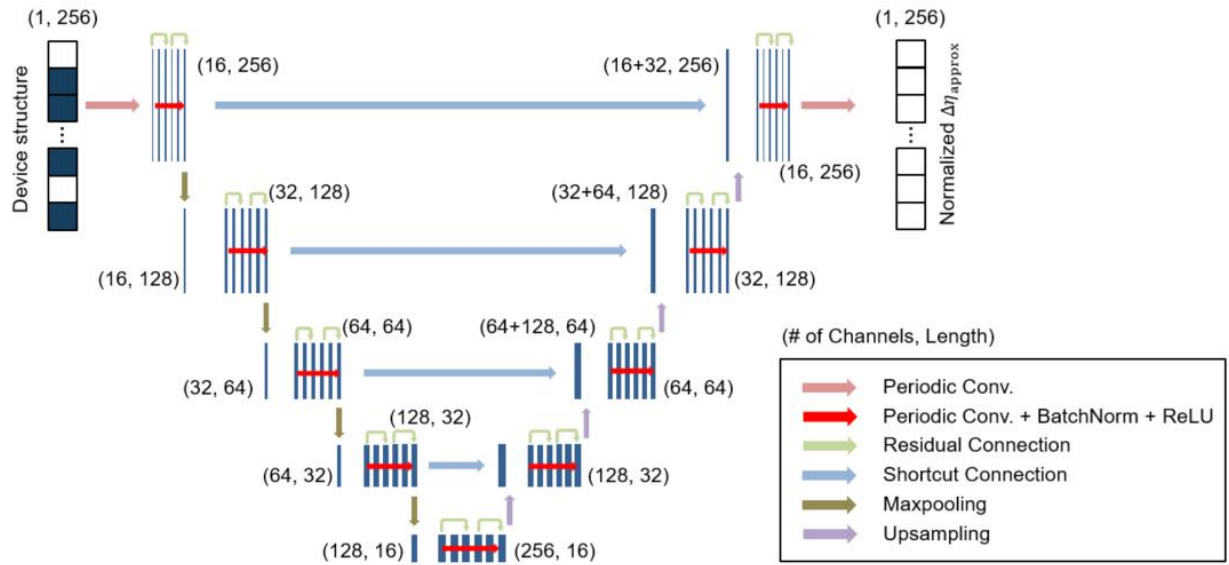

**Figure S1. Neural network architecture of the agent.** Number of channels and vector sizes are indicated as tuples.

## S2. Training dataset configuration

In the pre-training stage of PIRL, we use a dataset consisting of 20,000 structure-adjoint gradient pairs. The number of training dataset is selected based on the tradeoff between the prediction accuracy and the sample efficiency. We run five trials of supervised learning with each training data set consisting of 2,500, 5,000, 10,000, 20,000 and 40,000 samples. The calculated root-mean-square error (RMSE) is normalized with the standard deviation of the adjoint gradients in the training set. Figure S2a shows how the prediction error of the neural network changes as a function of the number of training samples available to the network.

The periodic nature of the system enables data augmentation through transversely displaced samples. The data-augmented training dataset has a size of 640,000. Also, as the samples with exceptionally high adjoint gradients obstruct the supervised learning process, they were excluded from the training dataset.

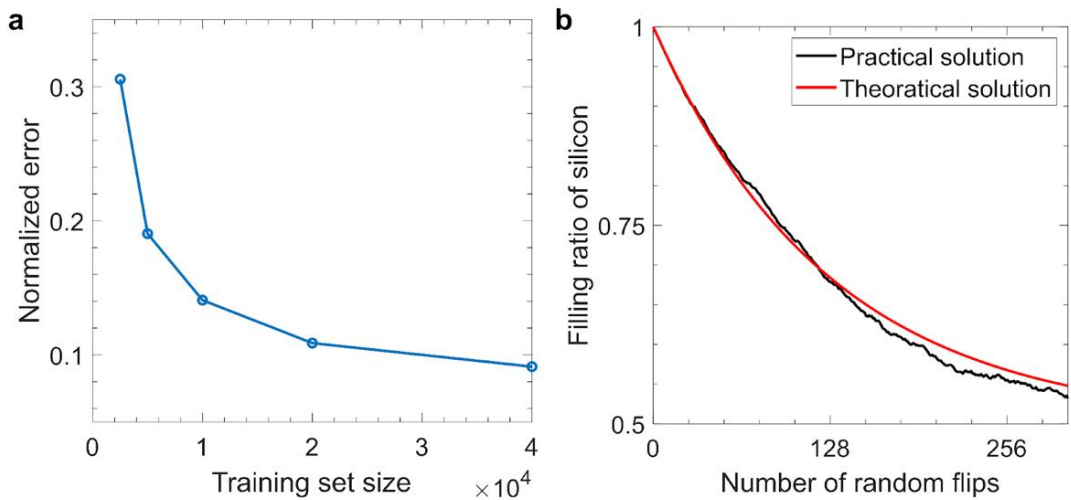

**Figure S2. a, The prediction error of the neural network for different training set sizes under a target condition λ = 1100 nm, θ = 60°.** Normalized error is calculated by dividing the test set root-mean-square-error (RMSE) by the standard deviation of the training set, ~0.0625. **b, Filling ratio of the Si cells during the random flipping.** Practical solution is estimated based on the Monte Carlo simulation and the theoretical solution is obtained through solving the differential equation $y' = [-y + (1 - y)]/256$, with initial condition $y(0) = 1$, where y represents a filling ratio of silicon cells during the random flipping of cells.

Randomly flipped samples are used to create the dataset for the pretraining stage. We plot the ratio of cells filled with silicon among the 256 cells assuming a full exploration (a fully-random flip). Each episode starts off with a silicon-filled device represented as [1, 1, …, 1]. Figure S2b shows the filling ratio of silicon as a function of the number of random flips. As the number of random flips increases, the filling ratio of the silicon converges to 50%. The silicon filling ratio of samples in the training dataset was set to follow the distribution given in Figure S2b as RL initially starts off from exploration dominated processes through Epsilon scheduling.

## S3. Comparison between the flip-value and the adjoint gradient

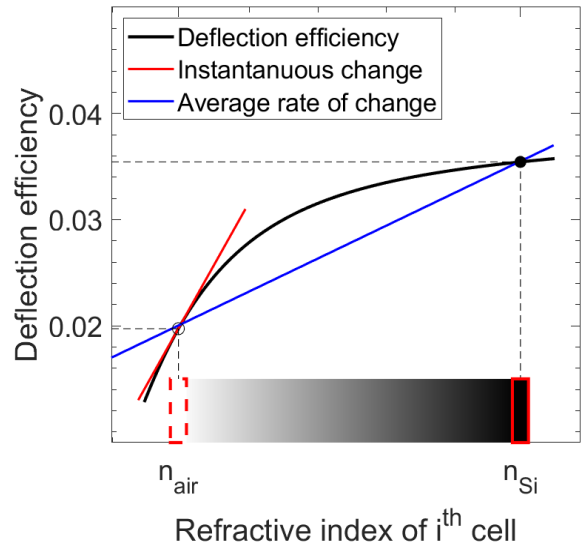

**Figure S3. The conceptual explanation of the adjoint gradient and the flip-difference calculation.** The deflection efficiency is plotted as a black line in the figure. The adjoint gradient is a local approximation (red) of the average rate of change in the flip-difference calculation (blue).

# S4. Algorithm summary

**Table S1. Pre-training stage**

---
**Algorithm 1** Pre-training stage of PIRL
---
**Require:** $\Delta\eta_{\text{approx}}$ calculator $g : s \rightarrow (\frac{\delta\eta}{\delta n}) \cdot \Delta n$, $rand_{flip} :=$ random flipping of a cell.
**Input:** Initial structure $s_{\text{Silicon}} = \{-1, +1, ..., +1\}$ and $s_{\text{Air}} = \{+1, -1, ..., -1\}$, training set size $N_{sample}$, maximum number of random flipping $N_{flip}$, approximator $f$, network parameter $\theta$.
**Output:** Training samples with size $N_{sample}$, optimized network parameters $\theta$.
1: **Initialization** Current device structure $s_0$, network parameters $\omega$.
2: **for** $1, 2, ..., (N_{sample}/2 \cdot N_{flip})$ **do**
3:    **for** $j = 1, 2, ..., N_{flip}$ **do**
4:      Set initial structure $s_0 = s_{\text{Silicon}}$
5:      **for** number of random walk $k = 0, 1, ..., j-1$ **do**
6:        Take action $a_k = rand_{flip}$ on structure $s_k$ and obtain $s_{k+1}$
7:      **end for**
8:      Calculate $\Delta\tilde{\eta}_{\text{approx}} = g(s_j)/\|g(s_j)\|_2$
9:      Store $(s_j, \Delta\tilde{\eta}_{\text{approx}})$ in training set
10:     Set initial structure $s_0 = s_{\text{Air}}$
11:     **for** number of random walk $k = 0, 1, ..., j-1$ **do**
12:       Take action $a_k : rand_{flip}$ on structure $s_k$ and obtain $s_{k+1}$
13:     **end for**
14:     Calculate $\Delta\tilde{\eta}_{\text{approx}} = g(s_j)/\|g(s_j)\|_2$
15:     Store $(s_j, \Delta\tilde{\eta}_{\text{approx}})$ in training set
16:   **end for**
17: **end for**
18: Pre-process samples in the training set
19: **if** *training stage* **then**
20:   Randomly sample $(s, \Delta\tilde{\eta}_{\text{approx}})$ from training set
21:   Calculate loss $L = \text{MSE}(\Delta\tilde{\eta}_{\text{approx}}, f_\theta(s))$
22:   Update network parameters $\theta \leftarrow \text{Adam}(\theta)$
23: **end if**
---

**Table S2. RL stage**

**Algorithm 2** Agent-environment interaction in a single worker
---
**Require:** Deflection efficiency calculator $f : s \rightarrow \eta$, $rand_{flip} :=$ random flipping of a cell.
**Input:** Initial structure $s_0 = \{-1, +1, ..., +1\}$, number of episodes $k$, network parameters $\omega$, target network parameters $\omega^-$, learning rate $\alpha$, discount factor $\gamma$.
**Output:** Trajectory $(s, a, r, s')$, optimized network parameters $\omega$.
1: **Initialization** Current device structure $s_0$, target network parameters $\omega' \leftarrow \omega$.
2: **for** number of episodes $i = 0, 1, ..., k-1$ **do**
3:    Set current structure $s_i$
4:    Take action $a_i : \begin{cases} \text{argmax}_a Q^\omega(s_i, a), & \text{with probability } 1-\epsilon \\ rand_{flip}(s_i), & \text{with probability } \epsilon \end{cases}$
   on structure $s_i$ and obtain $s_{i+1}$
5:    Calculate reward $r_i = \Delta\eta = \eta_{i+1} - \eta_i$, where $\eta_{i+1} = f(s_{i+1})$
6:    Store transition $(s_i, a_i, r_i, s_{i+1})$ in experience replay
7:    **if** *training stage* **then**
8:       Randomly sample transitions $(s_t, a_t, r_t, s_{t+1})$ from experience replay
9:       Calculate loss $L = \text{Huber}(y_t, Q^\omega(s, a))$,
      where Bellman target $y_t = r_t + \gamma \max_a Q^{\omega^-}(s_{t+1}, a_{t+1})$
10:      Update network parameters $\omega \leftarrow \text{Adam}(\omega)$
11:    **end if**
12:    **if** *update target network* **then**
13:      $\omega^- \leftarrow \omega$
14:    **end if**
15: **end for**

**Table S3. Genetic algorithm**

| Variable name | Value |
| --- | --- |
| MaxGeneration | 400 |
| PopulationSize | 500 |

We used the genetic algorithm module provided in the Global Optimization Toolbox in MATLAB. Table S4c outlines the value of hyperparameters used. Other hyperparameters are unchanged from their default values. Outline of each variable can be found in the following link: [1]

**Table S4. Greedy algorithm**

---
**Algorithm 3** Greedy depth 1 algorithm
---
**Require:** Deflection efficiency calculator $f : s \rightarrow \eta$.
**Input:** Randomly selected initial structure $s_0 \in [-1, 1]^N$ where $N$ is number of cells, action of flipping a cell in $i^{th}$ index $a(i)$, maximum number of iterations $k$. *tolerance* determines early stop condition.
**Output:** Best device structure $s_{\text{best}}$, and its efficiency $\eta_{\text{max}} = f(s_{\text{best}})$ after $256 \times k$ simulations.
 1: Initialize $s_{\text{best}} = s_0$, $\eta_{\text{max}} = 0$, $stop = 0$
 2: **while** $i \leq k$ **do**
 3:    Set current structure $s_i = s_{\text{best}}$
 4:    **for** cell index $j = 0, 1, ..., N - 1$ **do** {search for the best of all possible actions}
 5:       Take $a(j)$ on structure $s_i$ and obtain $s_i^{(j)}$
 6:       Calculate deflection efficiency $\eta^{(j)} = f(s_i^{(j)})$
 7:       Collect structure $s_i^{(j)}$ and efficiency $\eta^{(j)}$
 8:    **end for**
 9:    Take action $a_{\text{opt}}(k)$ where $k = \text{argmax}_j \eta^{(j)}$ and obtain $s_i^{a_{\text{opt}}}$, $\eta^{a_{\text{opt}}}$
10:    **if** $\eta_{\text{max}} < \eta^{a_{\text{opt}}}$ **then**
11:       Update best structure $s_{\text{best}} = s_i^{a_{\text{opt}}}$ and max efficiency $\eta_{\text{max}} = \eta^{a_{\text{opt}}}$
12:       $stop = 0$
13:    **end if**
14:    $i = i + 1$
15:    $stop = stop + 1$
16:    **if** $stop \geq tolerance$ **then**
17:       break
18:    **end if**
19: **end while**

---

**Table S5. Random search**

---
**Algorithm 4** Random search algorithm
---
**Require:** Deflection efficiency calculator $f : s \rightarrow \eta$, $rand :=$ a random device generator.
**Input:** Randomly selected initial structure $s_0 \in [-1, 1]^N$ where $N$ is number of cells, maximum number of iterations $k$.
**Output:** Best device structure $s_{\text{best}}$, and its efficiency $\eta_{\text{max}} = f(s_{\text{best}})$ after $k$ simulations.
 1: Initialize $s_{\text{best}} = s_0$, $\eta_{\text{max}} = 0$
 2: **while** $i \leq k$ **do**
 3:    Obtain random structure $s_i = rand$
 4:    Calculate deflection efficiency $\eta = f(s_i)$
 5:    **if** $\eta_{\text{max}} < \eta$ **then**
 6:       Update best structure $s_{\text{best}} = s_i$ and max efficiency $\eta_{\text{max}} = \eta$
 7:    **end if**
 8:    $i = i + 1$
 9: **end while**

---

We set the *tolerance* parameter $= 5$.

## S5. Hyperparameter table of reinforcement learning

All hyperparameters used in the RL stage are provided in Table S5. Description of each hyperparameter can be found in the RLlib website [2].

**Table S6. List of hyperparameters in RL**

| Variable name | Value |
| --- | --- |
| overall learning steps | $2\times10^5$ |
| learning_starts | 1,000 |
| target_network_update_freq | 2,000 |
| buffer capacity | $10^5$ |
| learning rate | 1000 |
| gamma | 0.99 |
| training batch size | 512 |
| number of training | 1 |
| horizon | 512 |
| number of rollout workers | 16 |
| number of environments per workers | 1 |
| initial epsilon | 0.99 |
| final epsilon | 0.01 |
| epsilon timesteps | 100,000 |

## S6. E-field distribution of an optimized device at a different target condition

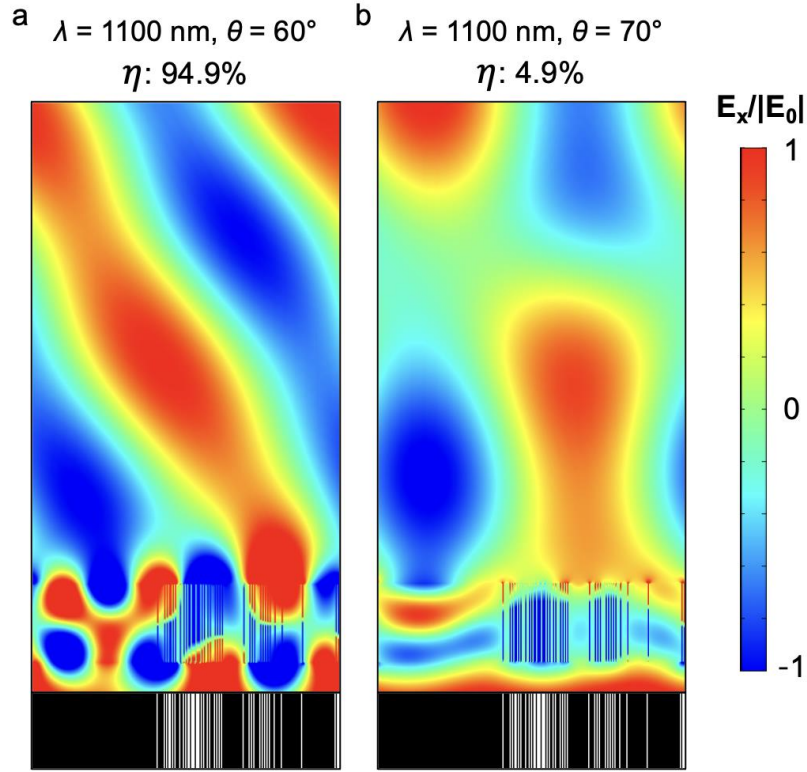

**Figure S4. Electric field distribution from an optimized beam deflector at (1100 nm, 60°) operating at a different target condition (1100 nm, 70°). a,** Device with the same state representation in Figure 4a. **b,** The structure shows deflection efficiency of 52.3% at (1100 nm, 70°) whereas the deflection efficiency at the original condition is 94.9%.

## S7. Transfer learning results: Mean and standard deviation

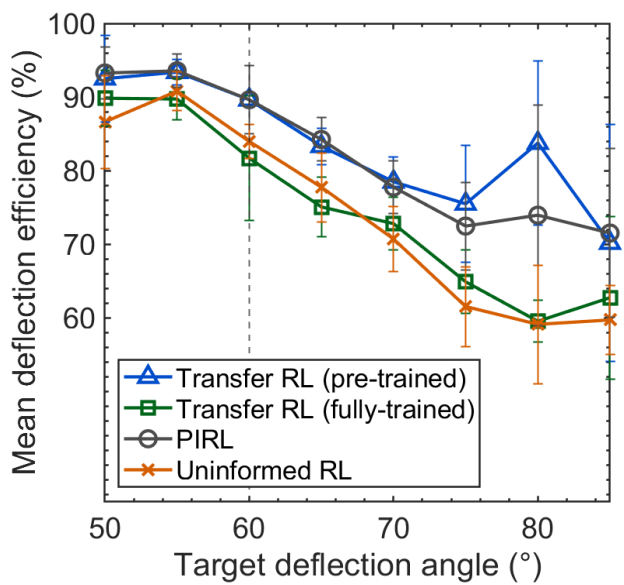

**Figure S5. The average deflection efficiency of the device found using PIRL (gray), uninformed RL (orange), and two different transfer learning processes: pre-trained model transfer RL (blue), and fully-trained model transfer RL (green).** The data points and error bar represent the average and standard deviation of maximum efficiencies from ten runs, respectively.